\documentstyle[12pt,aaspp4]{article}
\def\roma#1{\ifmmode{#1}\else{$#1$}\fi}

\def\kmsmpc{\roma{\rm\,km\,s^{-1}\,Mpc^{-1}}}                       
\def\Ho{\roma{\,\rm H_{o}}}                          
\def\qo{\roma{\,\rm q_{o}}}
\slugcomment{Submitted to PASP}

\begin{document}

\title{Imaging With STIS: Astronomy at V = 30}

\author{Michael D. Gregg$^{1}$ \& Dante Minniti$^{1}$}
\affil{Institute for Geophysics and Planetary Physics, L-413\\
Lawrence Livermore National Laboratory, Livermore, CA 94550\\
E-mail: gregg, dminniti@llnl.gov}

\begin{abstract}

In February, 1997, the second Space Shuttle servicing mission to the
Hubble Observatory will install the Space Telescope Imaging
Spectrograph (STIS).  This new instrument will greatly enhance the
spectroscopic capabilities of the Hubble Space Telescope by providing
a longslit format and CCD detector technology.  STIS can also be used
as an imager, providing an alternative to the Wide Field Planetary
Camera 2.  The filter set of STIS is limited and does not contain
standard bandpasses, but we show here that this does not preclude
useful broad band photometry.  In fact, the STIS photometric system
may be preferable to that of WFPC2 for certain applications where a
faint limiting magnitude and fine spatial resolution are overriding
considerations.

The two optical wide-band choices on STIS are a clear aperture and a
longpass ($\lambda > 5500$\AA) filter.  We define an effective
shortpass filter from the difference of these, making two-color
photometry possible with STIS.  We present preliminary transformations
between the STIS system and Cousins BVRI bandpasses, showing that
these transformations are very well-behaved over almost all
temperatures, luminosities, and abundances for normal stars.  In an
8-orbit cycle, STIS will be able to reach signal-to-noise of $\sim
5-10$ at V = 30.0 in its clear and longpass imaging modes, a
significant increase in the power of HST to address a number of
fundamental issues out of reach of current instrumentation
capabilities on the ground or in space.

\end{abstract}


\section{Introduction}


With its combination of high angular resolution and modern CCDs, the
Hubble Space Telescope has recently achieved the new world record for
deep imaging in the Hubble Deep Field project (Williams et al.\ 1996).
By integrating for $\sim 100,000$ seconds with the Wide Field
Planetary Camera 2 (WFPC2), objects at 30$^{th}$ magnitude in the
F606W (roughly V) filter have been detected with S/N of 3 to 4.  The
second Space Shuttle servicing mission to HST is scheduled for early
in 1997.  One of the new instruments to be placed aboard HST is the
Space Telescope Imaging Spectrograph (STIS).  Although it is much
anticipated as a two-dimensional spectrograph, its imaging
capabilities will be extremely useful.  In this paper, we explore the
photometric possibilities of STIS, deriving a synthetic calibration.
We argue that where a faint limiting magnitude is the overriding
criterion, STIS will be superior to WFPC2.  Even though STIS has a
very limited filter set in the optical, we show that it can be
exploited to provide useful broad band color information to
unprecedented depth.  Even considering the August 1996 announcement
that throughput will be 30\% below the original expectations, STIS is
still expected to reach $\sim 1.5$ magnitudes fainter than WFPC2 for a
given integration time.  This brings the solutions to a range of
heretofore unaddressable problems within reach and holds out the
possibility that astronomy at V = 30 will become regular with HST.

\section{Definition of the STIS Wide-Band Photometric System}

The STIS CCD is a SITe 1024x1024 thinned CCD with $21\mu$ pixels and a
16 bit analog-to-digital converter.  The focal plane scale is
0\farcs05 per pixel.  The FWHM is close to 2 pixels at 5000\AA, and
the 90\% encircled energy radius is 3 pixels.  The STIS Instrument
Handbook notes that the point spread function is expected to degrade by 30\% at the
extreme top and bottom of the field of view.  Complete details of STIS
as an imager can be found in the STIS Instrument Handbook (Baum et
al.\ 1996).

The STIS imaging filter set includes a clear aperture, officially
named 50CCD, with a field of view of 50\arcsec x 50\arcsec.  Also
available is a longpass ($\lambda > 5500$\AA) filter, officially named
F28x50LP, which has a field of view of 28\arcsec x 50\arcsec.  We
refer to these two bands simply as CL and LP, respectively.  The
response curve of the CL bandpass is set entirely by the telescope
throughput and CCD sensitivity.  The response of the LP bandpass is
very similar to the CL, only slightly lower, at all wavelengths longer
than 5500\AA.  Below this wavelength, the LP throughput is essentially
zero (Figure~1; see also Figure 5.3 in the STIS Instrument Handbook).

The close match of the two bandpasses longward of 5500\AA\ is
fortuitous because it allows an effective shortpass (SP) filter and
(SP-LP) color to be
defined based on differencing the CL and LP bandpasses.  The STIS
broad band photometric system is then given by
\begin{eqnarray}
CL & = & -2.5log({F_{CL}}) + C_{CL},\\
LP & = & -2.5log({F_{LP}}) + C_{LP},\\
SP & = & -2.5log({F_{CL} - F_{LP}}) + C_{SP},~~~ {\rm and} \\
SP - LP & = & -2.5log(\case{F_{CL} - F_{LP}}{F_{LP}}) + C_{SPLP}.
\end{eqnarray}
where $F_{CL}$ and $F_{LP}$ are the total counts for an object in each
bandpass, and the $C_i$ are zeropoint constants.  This very wide-band
system resembles to some extent that used successfully by the MACHO
collaboration to discover and monitor microlensing events (e.g. Alcock
et al.\ 1995).  The MACHO filter system has been calibrated on to
standard systems and has produced accurate photometry and
color-magnitude diagrams.  The slight mismatch between the CL and LP
bandpasses (Figure~1) means that the SP bandpass has a small red leak,
but in what follows, this does not appear to be significant except for
the very reddest stars.

\section{Synthetic Calibration of STIS Photometry}

To investigate the utility of the STIS photometric system, we have
derived transformations between the STIS magnitudes and (SP-LP) color
and the Cousins wide-band system by convolving the Bruzual et al.\
spectrophotometric stellar library with the sensitivity curves
(normalized) for the CL and LP bandpasses that are given in Chapter 14
of the STIS Instrument Handbook.  We have set the zeropoint using the
Kurucz (1992) model for Vega.  This procedure is similar to that used
by Holtzman et al.\ (1995) in calibrating the WFPC2 photometric
system.  The transformations between (SP-LP) and many of the standard
broad band colors are extremely well-behaved over much of the color
range, especially (B-R), (B-I) and (V-R) (Figure~2).  Stars of
luminosity class V are indicated by the symbols with black dots.  The
solid lines in Figure~2 are cubic spline fits obtained interactively
using the task CURFIT in IRAF and the RMS about each fit, excluding
obvious outliers, is shown in the figure.  The dotted lines have a
slope of unity and are drawn for reference.

The residuals from the spline fits are displayed for each color.
There is a small but clear systematic separation between dwarfs and
giants for spectral types later than K5 ($SP-LP > 1.5$), but it is at
a level of 0.1 magnitudes or less.  This behavior is seen in
transformations in other photometric systems and may even be
advantageous in some circumstances.  Quantifying this refinement is
beyond the scope of this analysis, but if the late dwarfs and giants
do separate cleanly, then there are apparently a few errors in the
luminosity classifications in the Bruzual et al.\ stellar library.

The most natural transformation is between (SP-LP) and (B-R), which
are almost perfectly matched over the interval ($0<B-R<2$), where the
scatter is only $\pm 0.02$ mag.  The increase in scatter for $(SP-LP)
\gtrsim 2$ may be a consequence of the red leak in the SP bandpass.
The transformation between (SP-LP) and (B-I), while not having a slope
of 1, is well-behaved over an even larger color range.

The cubic spline fits make it difficult to supply transformation
equations in a concise manner.  Instead, a look-up table is given in
steps of 0.1 magnitudes in $(SP-LP)$ for converting to Cousins broad
band colors (Table~1).  An electronic version of this table, in steps
of 0.01 magnitudes, is available from the authors.

In Figure~3, we plot comparisons of the difference between SP, CL, and
LP magnitudes with Cousins B, V, R, and I as a function of $(SP-LP)$.
The lines are cubic spline fits and, as above, the RMS is given in the
figure.  Again, the comparison shows that well-behaved transformations
exist over almost all spectral types.  For stars earlier than mid-M,
the RMS scatter is only $\sim 0.02$ magnitudes in all bands.  Here
too, rather than supply a transformation equation, we simply list in
Table~1 the differences between the STIS magnitudes and Cousins B, V,
R, and I as given by the spline fits plotted in Figure~3.

The Bruzual et al.\ stellar library contains spectra of stars mainly
of solar metallicity and luminosity class V and III covering all
spectral types.  Figures 2 and 3 demonstrate that there is a small
separation by luminosity class in the STIS colors, of a few hundredths
to a tenth of a magnitude, mainly for spectral types later than K5.
We have also tested for possible metallicity effects using the model
spectral energy distributions of Kurucz (1992).  Figure~4 shows
$(B-R)$ and $(SP-LP)$ for models covering the the temperature range
3500 to 50000 K at two extreme abundances, [M/H]$ = -2.0$ (crosses)
and +0.5 (triangles).  There is a small difference for stars with $1 <
(SP-LP) < 2$ at the level of a few hundredths of a magnitude.  For the
very reddest models, there is a larger difference in the
transformation, but this may be a shortcoming of the models.  The
solid line in Figure~4 is a spline fit to the data shown in Figure~2;
this fit clearly tracks the [M/H]$ = -2.0$ models better than the
higher metallicity ones, even though the observed stars are
approximately solar abundance.  The spline fit to the observed stars
also differs systematically at the blue end, but here too only at the
few hundredths of a magnitude level, perhaps also indicating minor
difficulties with the theoretical spectral energy distributions.

These derived zeropoints and transformations do not completely
substitute for an empirical calibration of the STIS passbands on board HST
when precise sensitivity functions and throughput can be determined,
but they do demonstrate that STIS is useful for broad band photometry
and that well-determined transformations to the standard Cousins
passbands exist.  Given the well-determined and calibrated
transformations between the Cousins system and that of WFPC2 (Holtzman
et al.\ 1995), there is no doubt that transformations between STIS and
WFPC2 will also be equally well-behaved.  This raises the possibility
of using the unprecedented deep imaging capability and spatial
resolution of STIS for a number of problems that cannot be addressed
with current instrumentation.

One of the most important applications of broad band photometry is to
construct color-magnitude diagrams for stellar systems to derive ages
and abundances.  The transformations derived here make it possible
to place isochrones calculated in Cousins bandpasses on the
STIS system, as Holtzman et al.\ did for the WFPC2 filters.
We have mapped the New Yale Isochrones (Demarque et al.\ 1996) to the STIS
photometric system.  As examples, Figure~5 shows solar and 1/100 solar
metallicity isochrones for 5, 10, and 15 Gyrs in the STIS CL vs.\
(SP-LP) plane.  Because of the extremely well-behaved transformations,
the isochrones in the STIS system are completely normal in appearance,
their utility undiminished.

\section{STIS Imaging Exposure Time Calculator}

We have written a program to make careful estimates of the expected
S/N for STIS in imaging mode.  The program convolves spectrophotometry
of representative stars from the Bruzual et al.\ library with the STIS
CL and LP sensitivity functions, summing over each bandpass to obtain
a count rate.  Given the known instrumental characteristics of STIS,
the S/N can be estimated using equations from the STIS Handbook, pp.\
70-71.  We have reduced the listed sensitivities by 30\% as advised in
the August STIS Instrument Handbook update.

Our estimates indicate that, given a low sky background and working in
relatively uncrowded fields, a S/N level of 5 can be reached for a
total exposure time of approximately 16000 seconds in the CL band and
21500 seconds in LP for a G5V star with V = 30.0.  The S/N achieved in
SP will be lower because the errors in both CL and LP contribute in
quadrature.  The color $(SP-LP)$ is doubly sensitive to the error in
LP (Equation 4).  We have taken into account that CL exposures will be
limited to 900 seconds to avoid the possibility of UV damage to the
CCD (STIS Instrument Handbook).  The extra noise from the frequent
reads has an impact on the CL S/N and has the effect of nearly
equalizing the exposure times between the two bands for longer total
exposures to reach a given S/N.  Although these exposure times are
lengthy, they can be accommodated in two 8-orbit intervals between
passages of HST through the South Atlantic Anomaly.

This faint magnitude limit reached by STIS exceeds that of WFPC2
because of several contributing factors: 1) the STIS CCD has 1.5 times
the quantum efficiency of the WFPC2 CCDs, 2) the STIS CCD has lower
readout noise, and 3) the STIS passbands are much broader than any of
the WFPC2 filters.  Compared to the imaging S/N calculation examples
in the STIS Handbook, our detailed estimates are conservative by $\sim
10-15\%$, even after taking into account the reduction in sensitivity
described in the August 1996 Update.  This is likely due to adopting a
more conservative point spread function for calculating the fractional
energy received within the measured aperture.  This may be appropriate
because of the two additional sources of noise described in the August
1996 update to the STIS Instrument Handbook of fringing at $\lambda >
7000$\AA, and "halos" around red objects at the $10^{-3}$ to $10^{-4}$
level.  The full ramifications of these effects on S/N will not be
understood until STIS is calibrated and evaluated on board HST.

We stress the need for a space-based empirical calibration of STIS
photometry.  Because of the unusually wide wavelength coverage of the
STIS passbands and the sensitivity down to well below the atmospheric
cutoff in the CL band, the STIS system will be extremely difficult to
calibrate reliably from the ground.  An absolute calibration of the
STIS photometric system from space will be needed to take full
advantage of its potential.  Such calibrations can be done with a
small amount of HST time and will prove useful for a wide range of
projects which make use of STIS photometry.  One possibility is to
image a grid of fields in Galactic globular clusters which have
accurate ground-based Cousins BVRI and space-based WFPC2 wide filter
photometry.  The clusters should span as much range in metallicity as
possible to establish the exact color transforms and zeropoints.  Such
images would include thousands of stars each, which, combined with
archival WFPC2 images, would provide all the needed zeropoint and
transformation data and would even allow exploration of second order
metallicity and luminosity effects.

\section{Summary}

The advent of STIS imaging on HST promises to open new frontiers for
investigation which are beyond the reach of current instrumentation,
either in space or on the ground.  Although the bandpasses for optical
imaging on STIS are limited, the analysis presented here demonstrates
that meaningful two-color photometry is possible.  The unusually wide
bandpasses of STIS are in fact an advantage, allowing photometry to
unprecedented depth to be brought to bear on a number of fundamental
questions.

There are many areas where STIS imaging can have a significant impact.
We mention a few examples, which by no means cover all the fields
that could be impacted.  The reader's imagination is left to consider
further possibilities.

\noindent
$\bullet$ {\it Main Sequence Turnoff Throughout the Local Group}

The star formation history of Local Group galaxies is varied (Hodge
1989; Da Costa 1995).  With the new imaging capabilities of STIS, the
main sequence turnoff's of all Local Group galaxies are now within
reach, offering the possibility of directly probing the ages of
objects which collapsed from primordial clouds independent from the
Milky Way.  The deepest photometry to date in M31 is from WFPC2 (e.g.,
Fusi-Pecci et al.\ 1996; Holland, et al.\ 1996) which extends 1
magnitude below the horizontal branch.  The oldest main sequence
turnoff of the halo of M31 (true distance modulus 24.4; Freedman \&
Madore 1990), for example, will be at ${\rm V} \approx 29.0$ for a
$t\sim 15$ Gyr, solar abundance population (cf.\ Figure 5).  Deep STIS
exposures with the clear aperture and longpass filter will yield
6--12$\sigma$ magnitudes at $V = 29$, and would allow the turnoff to
be estimated to 0.1--0.2 mag with an age resolution of $\sim 2-3$ Gyr.
Knowing the turnoff ages for many Local Group galaxies would
effectively document the star formation history of a wide range of
galaxy types and probe directly the formation process of the Local
Group in particular and galaxies in general.

\noindent
$\bullet$ {\it Cepheids in the Coma Cluster}

The debate over the Hubble constant continues to rage (Kennicutt,
Freedman \& Mould 1995) with recent values ranging from $56-58 \pm 4$
\kmsmpc\ (Sandage et al.\ 1996) to $81 \pm 6$ \kmsmpc\ (Tonry et
al.\ 1996), differing at the $4\sigma$ level.  With STIS, it will be
possible to detect and phase Cepheids at cosmological distances where
the peculiar velocities of individual galaxies and clusters are
insignificant compared to the smooth Hubble flow.  There is consensus
in the literature on the Cepheid distance scale, so a measure of
Cepheid distances to spirals in clusters out to the distance of Coma
cluster would go a long way toward reducing the present uncertainty in
the Hubble constant.  STIS two-color photometry can be used
effectively to positively identify and phase Cepheids, and to measure
their reddening.  If the Hubble constant is high, 70-80 \kmsmpc, then
the Coma cluster is $\sim 90$ to 100 Mpc distant, corresponding to a
distance modulus of $\sim 34.8$ to 35.  Cepheid variables with periods
of 50-65 days have mean M$_{\rm V} = -6$ to $-6.3$.  For an 8 orbit
integration in CL, our exposure time calculator estimates that these
variables could be measured with S/N 6 to 13 at mean light.  This is
comparable to the S/N achieved for the shortest period (lowest
luminosity) Cepheids measured in Virgo cluster spirals (Ferraresce et
al.\ 1996).  For a number of Cepheids in even a single spiral in the
Coma cluster, such observations can produce a Hubble constant good to
10\% or better.

\noindent
$\bullet$ {\it Red Giant Branch Tip in Early Type Galaxies out to Virgo}

Stellar population studies of giant early type galaxies,
ellipticals in particular, are hampered by the lack of nearby examples
which can be studied star-by-star.  Several analogs are often used,
such as the bulges of spirals or dwarf ellipticals in the Local Group.
Our understanding of the evolution of early type systems is critical
for deriving \qo\ and, to some extent, even \Ho.  
STIS, with the help of the Near Infrared Camera and Multi-Object
Spectrograph (NICMOS, Axon et al.\ 1996), also scheduled to be installed in
the same HST servicing mission, can make a dramatic improvement in the
ability to probe directly the stellar populations in giant
ellipticals.  The tip of the red giant branch for old populations has
M$_{\rm I} \approx -4$ and M$_{\rm H} \approx -7$; to measure this
well, it is necessary to go at least 2 magnitudes fainter with some
precision.  Establishing the nature of the upper red giant branch
would set limits on the metallicity and metallicity spread, the
presence of intermediate age populations, and the distance to
early type galaxies.  
With 15 orbits of integration, color-magnitude diagrams for the upper
giant branch can be observed with STIS LP and NICMOS H in ellipticals
out to Virgo cluster distances ($\mu \approx 31.1$), offering the
chance to begin detailed investigation of the brightest stellar
populations in early type galaxies.

\noindent
$\bullet$ {\it Baryonic Dark Matter in Nearby Galaxies}

Velocity dispersions of nearby dwarf spheroidal galaxies are generally
larger than can be accounted for with normal stellar populations, and
it is believed that dwarf spheroidal galaxies contain substantial
amounts of dark matter (e.g.\ Aaronson 1983, Mateo et al. 1991).  Deep
STIS imaging could probe deep enough down the main sequence (down to a
few $\times 0.1~{\rm M}_{\odot}$) in the nearest dwarf spheroidal
galaxies such as Fornax and Ursa Minor, to test if low mass stars are
the culprits.  Deep luminosity functions in these systems will reveal
if the mass function is skewed towards low-mass stars, contributing to
our knowledge of the dark matter observed in much larger scales.  Even
if the mass in the lower main sequence of dwarf spheroidals is not
enough to account for their unusually large velocity dispersions,
establishing the mass function to such low levels has implications for
both star and galaxy formation.

\noindent
$\bullet$ {\it The Hubble Deep Field South}

The faint magnitude limit accessible with STIS would also allow a
repeat of the Hubble Deep Field (HDF) experience (Williams et al.\
1996) in a more economical manner in the Southern hemisphere.  Only 16
orbits total for both CL and LP in the continuous viewing zone would
be needed to achieve the same limiting magnitudes as the HDF -North-,
although limited to only the two bandpasses and smaller field of view
of STIS.  If a total of 150 orbits were also allocated to the HDF
South, then either the areal coverage could be increased over a single
STIS field or the magnitude limits could be pushed even fainter than
the HDF North, reaching S/N $\sim 5$ at V$ = 31.0$ for a 100,000
second integration.  Additional color information could be obtained
with parallel NICMOS observations, perhaps of the same pieces of sky
by dividing the time between two fields separated by the STIS - NICMOS
focal plane distance and flipping the telescope $180\arcdeg$ between
observations.

\acknowledgements 

We would like to thank our collaborators K. Cook, E. Olszewski,
M. Mateo, and G. Da Costa for guidance and discussions.  We
are grateful to the STIS Science team headed by S.~Baum and the
Help Desk at STScI headed by D.\ Golombek for their prompt replies to
our inquiries.

\pagebreak
\begin{figure}[t]
\plotfiddle{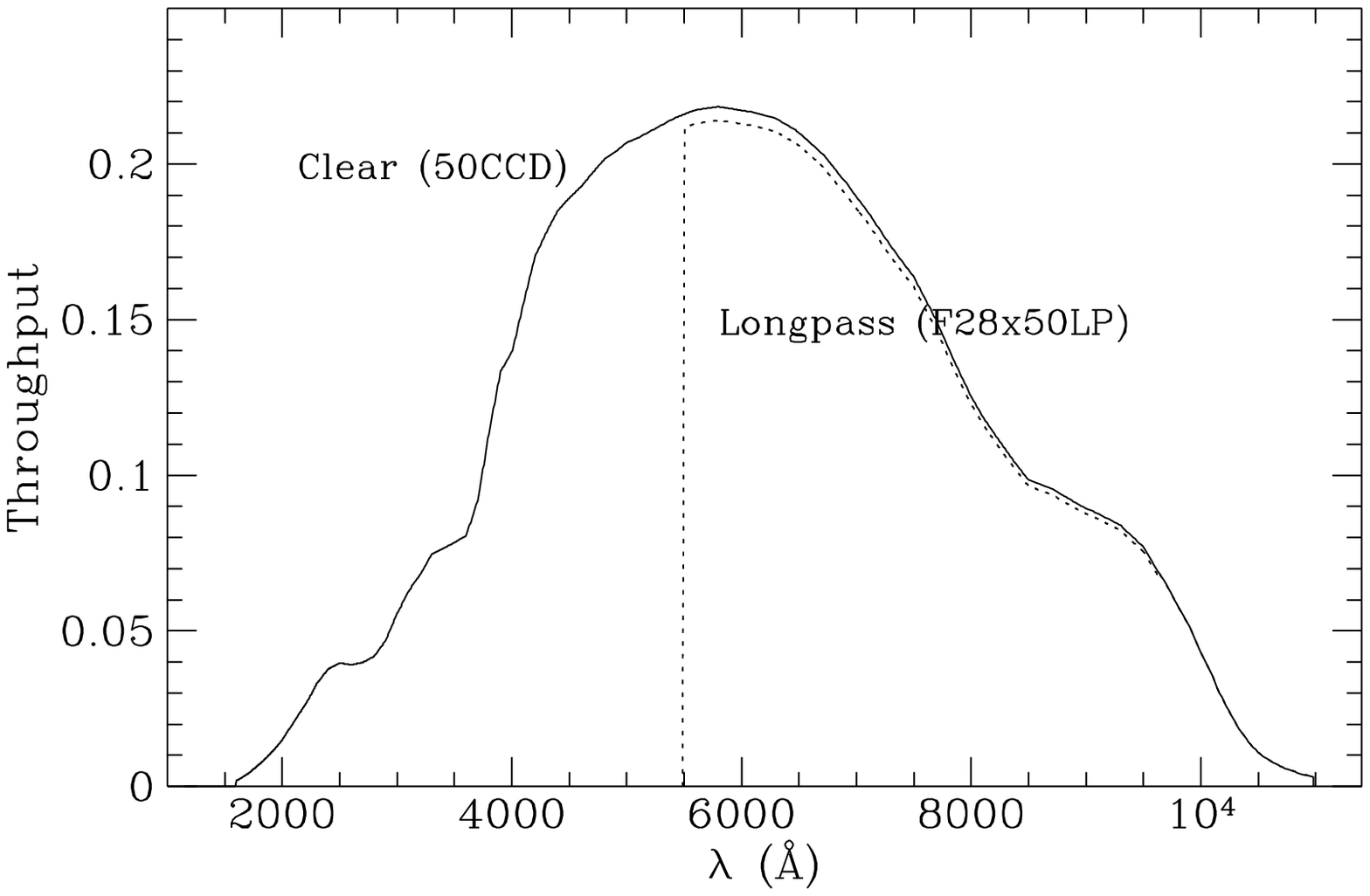}{3.5in}{0}{90}{90}{-260}{-265}
\caption{Throughput of the STIS wide bandpasses as given in the STIS
Instrument Handbook.  The Clear 50CCD (CL) curve is shown by the solid
line and the Long Pass F28x50CCD (LP) is the dotted line.  Note the
close similarity of the two everywhere the LP transmits.}
\end{figure}

\begin{figure}[p]
\plotfiddle{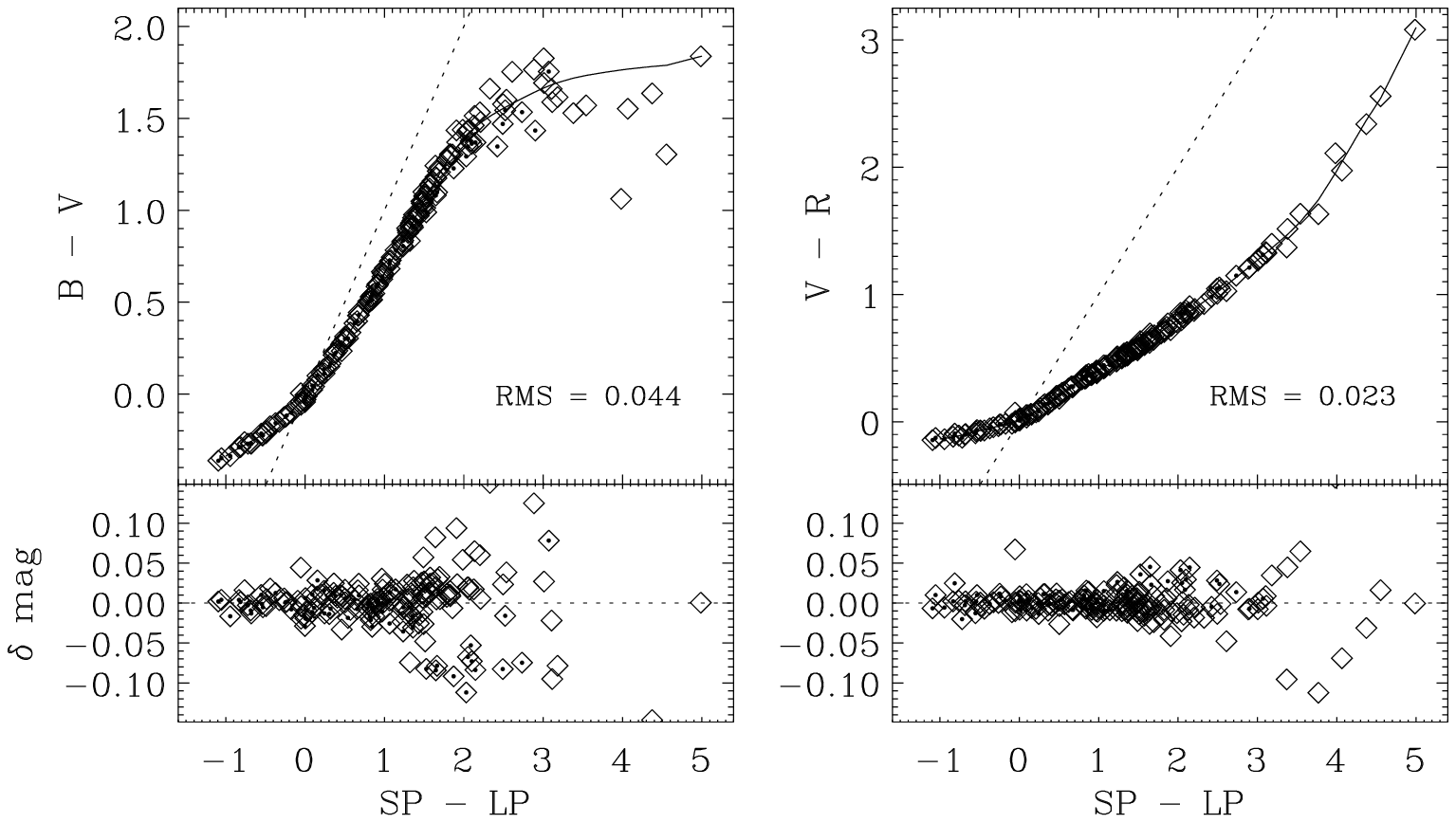}{3.3in}{0}{100}{100}{-300}{-250}
\plotfiddle{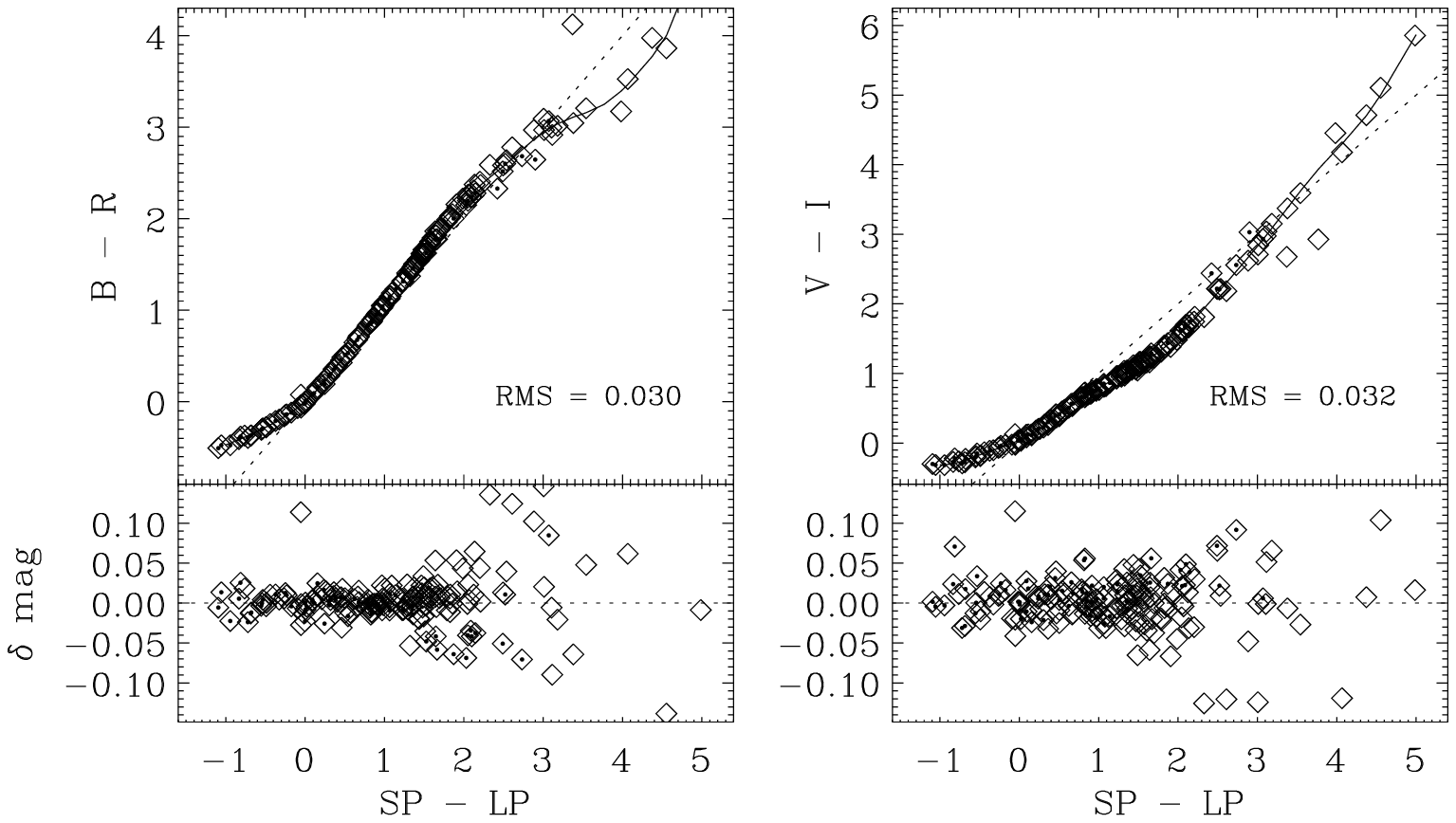}{3.3in}{0}{100}{100}{-300}{-265}
\caption[]{ Transformations of the STIS broad band color, $SP-LP$, to
standard Cousins filters, determined by convolving the STIS sensitivity
functions with the spectral energy distributions of 175 stars from the
spectrophotometric library of Bruzual et al.  Dwarfs are indicated by
the symbols containing black dots. The dashed lines
have a slope of unity.  The best transformations are between $SP-LP$
and $B-R$ or $B-I$.  There is a small systematic difference in the
transformations for stars later than spectral type K5 ($SP-LP >
1.5$), but it is at the few hundredths of a magnitude level.}
\end{figure}

\begin{figure}[t]
\figurenum{2}
\plotfiddle{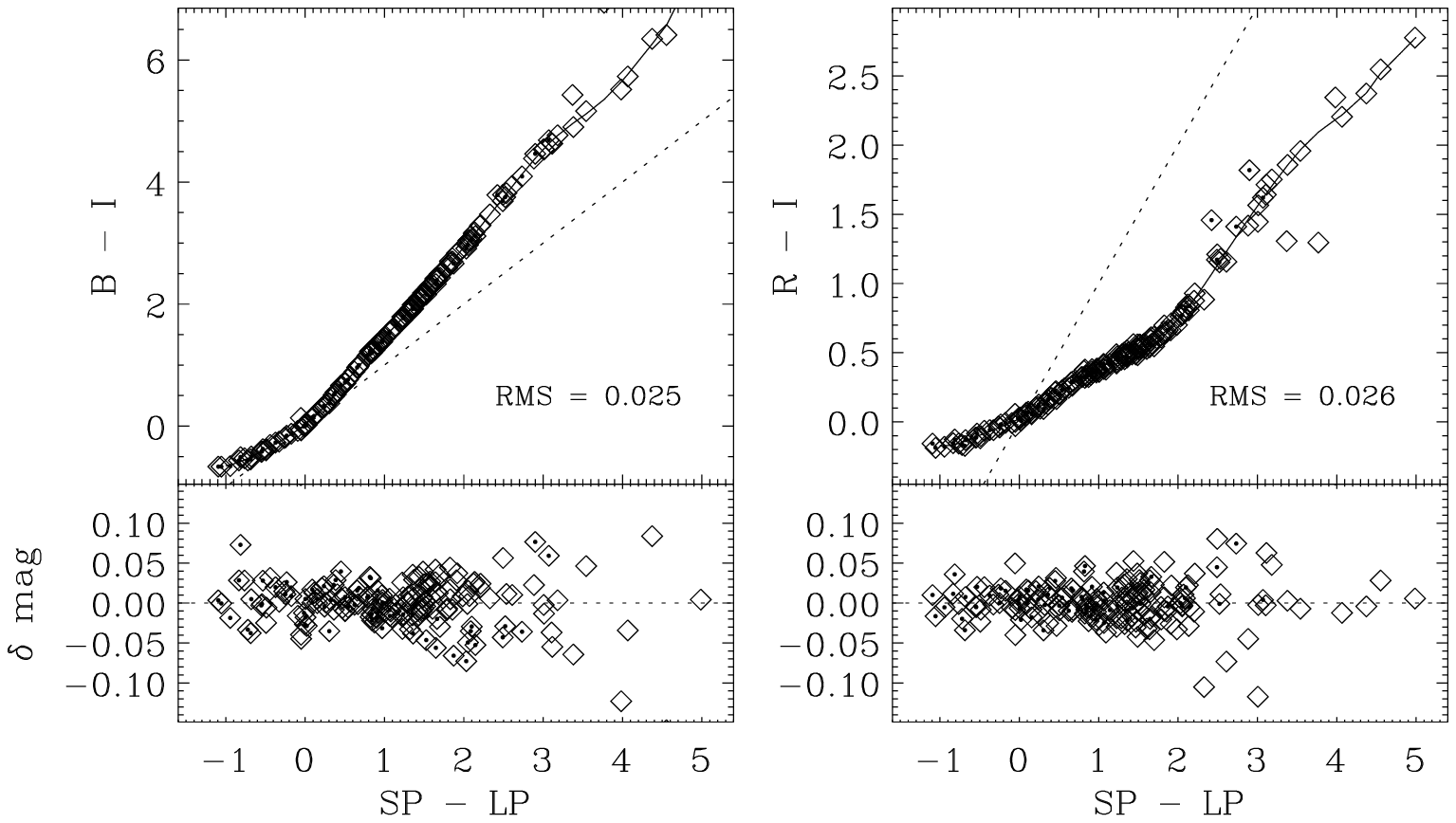}{3.7in}{0}{100}{100}{-300}{-260}
\caption[]{ Figure~2 cont'd}
\end{figure}

\begin{figure}[p]
\plotone{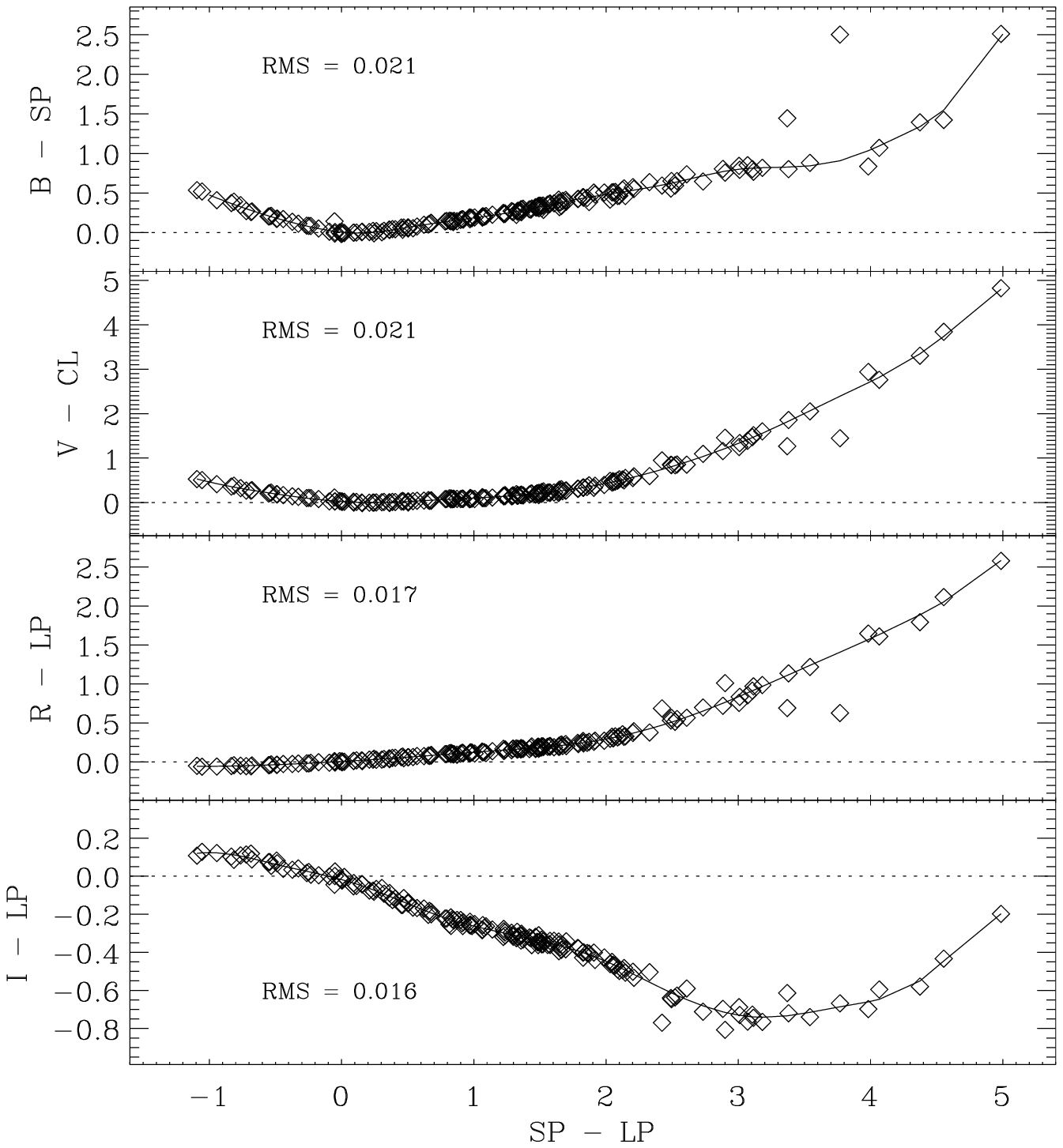}
\caption[]{ Transformation of the STIS band passes SP, CL, and LP to
standard Cousins B, V, R, and I, as a function of (SP--LP) determined
from the spectral energy distributions of 175 stars from the
spectrophotometric library of Bruzual et al.  The Kurucz
(1992) model of Vega is used to set the zeropoint }
\end{figure}

\begin{figure}[t]
\plotfiddle{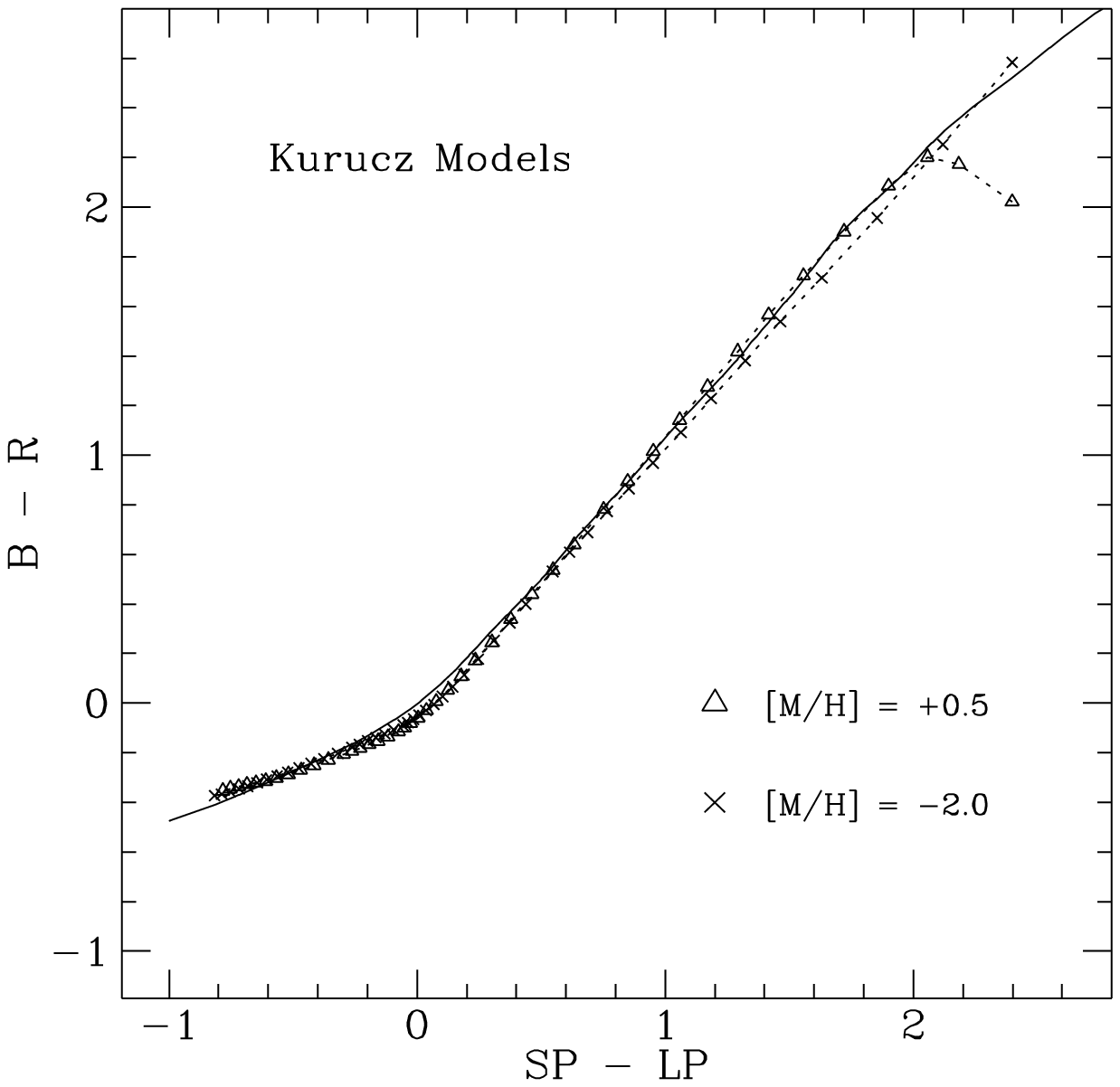}{3.7in}{0}{85}{85}{-260}{-200}
\caption[]{ Transformations of the STIS broad band color, $SP-LP$, to
standard Cousins filters, for Kurucz model atmospheres for two
abundances, [M/H]$ = -2.0$ and $+0.5$.  Any metallicity affects in the
transformation are very small, at the level of a few hundredths of a
magnitude.  The solid line is the same spline transformation plotted
in Figure~2 for stars in the Bruzual et al.\ spectrophotometric
library between $SP-LP$ and $B-R$.}
\end{figure}

\begin{figure}[t]
\plotfiddle{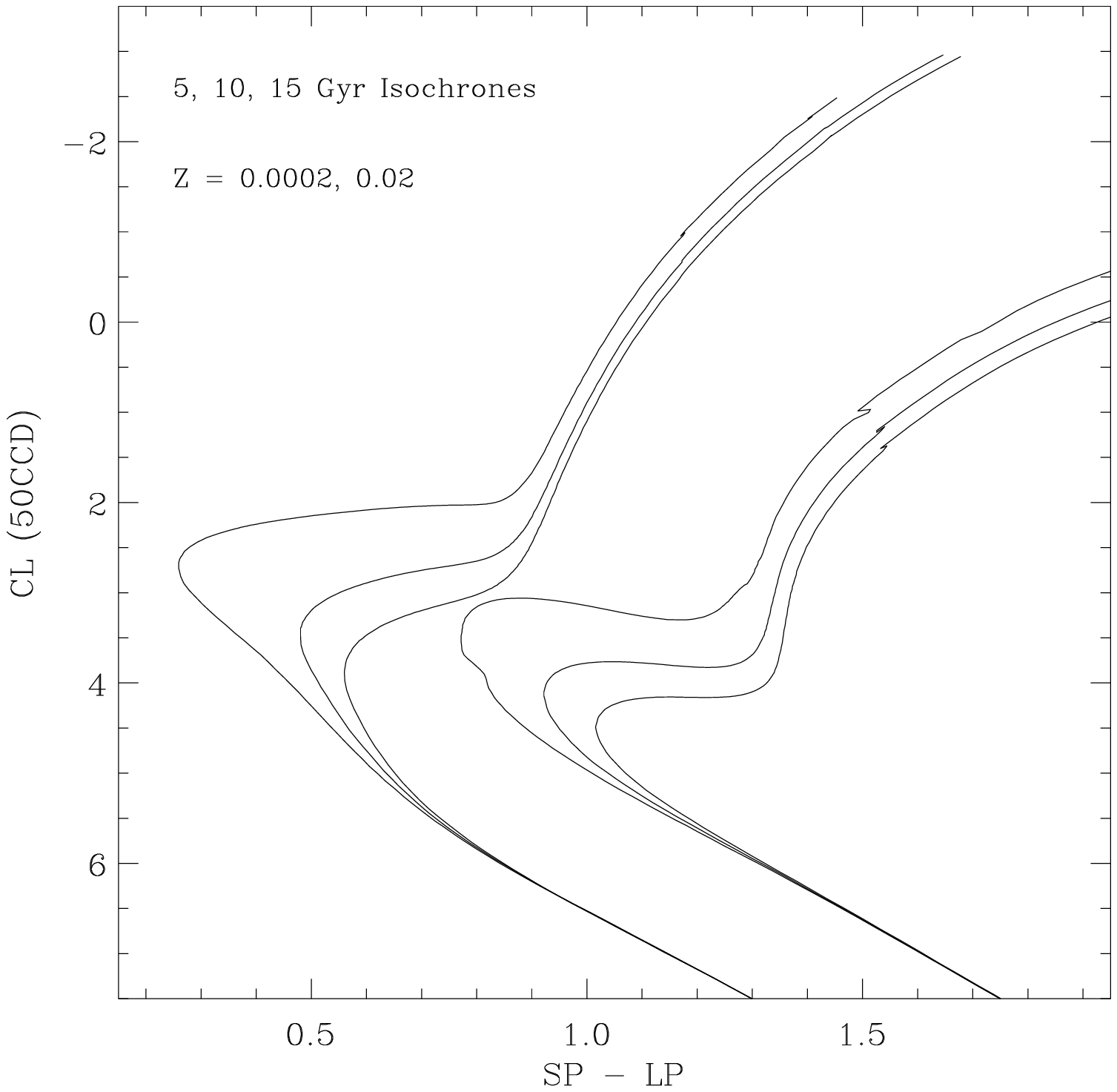}{4.1in}{0}{85}{85}{-240}{-135}
\caption[]{ Example New Yale Isochrones (Demarque et al.\ 1996)
transformed to the STIS photometric system.  Two limiting cases are
shown, metal-poor and metal-rich, for three different ages.  The
metal-poor turnoffs are bluer and brighter than the metal-rich
turnoffs of the same age.  Although the isochrones shown are
unreddened, a reddening of $E_{B-V} = 0.2$ has been assumed to compute
the S/N = 5 line drawn.  We have also computed the dependence of S/N
with color, and it is nearly flat as indicated.  }
\end{figure}

\begin{deluxetable}{rrrrrrrrrrr}
\footnotesize
\tablenum{1}
\tablewidth{0pt}
\tablecaption{STIS to Cousins Interpolation Table}
\tablehead {
\colhead {SP--LP} & 
\colhead {B--V} &
\colhead{V--R} & 
\colhead{B--R}   & 
\colhead{V--I} & 
\colhead{B--I}    & 
\colhead{R--I} &
\colhead{B--SP} &
\colhead{V--CL} &
\colhead{R--LP} &
\colhead{I--LP} 
} 
\startdata
  -1.000 & -0.338 & -0.134 & -0.471 & -0.312 & -0.650 & -0.173 & 0.475 & 0.463 &-0.054 & 0.123 \nl
  -0.900 & -0.311 & -0.127 & -0.437 & -0.307 & -0.618 & -0.171 & 0.410 & 0.397 &-0.053 & 0.119 \nl
  -0.800 & -0.286 & -0.116 & -0.401 & -0.287 & -0.573 & -0.161 & 0.348 & 0.339 &-0.050 & 0.110 \nl
  -0.700 & -0.262 & -0.102 & -0.363 & -0.253 & -0.516 & -0.143 & 0.290 & 0.286 &-0.046 & 0.096 \nl
  -0.600 & -0.237 & -0.086 & -0.322 & -0.212 & -0.449 & -0.121 & 0.237 & 0.240 &-0.041 & 0.080 \nl
  -0.500 & -0.211 & -0.070 & -0.280 & -0.170 & -0.380 & -0.098 & 0.186 & 0.195 &-0.035 & 0.063 \nl
  -0.400 & -0.180 & -0.055 & -0.234 & -0.130 & -0.309 & -0.075 & 0.140 & 0.152 &-0.027 & 0.047 \nl
  -0.300 & -0.145 & -0.040 & -0.184 & -0.091 & -0.235 & -0.052 & 0.098 & 0.113 &-0.020 & 0.031 \nl
  -0.200 & -0.105 & -0.024 & -0.129 & -0.051 & -0.156 & -0.029 & 0.060 & 0.077 &-0.011 & 0.016 \nl
  -0.100 & -0.062 & -0.005 & -0.068 & -0.007 & -0.069 & -0.004 & 0.029 & 0.046 &-0.003 &-0.000 \nl
   0.000 & -0.013 &  0.017 &  0.001 &  0.042 &  0.028 &  0.023 & 0.006 & 0.022 & 0.007 &-0.018 \nl
   0.100 &  0.040 &  0.045 &  0.082 &  0.099 &  0.138 &  0.053 &-0.004 & 0.007 & 0.017 &-0.038 \nl
   0.200 &  0.098 &  0.079 &  0.175 &  0.166 &  0.263 &  0.088 &-0.000 & 0.000 & 0.027 &-0.062 \nl
   0.300 &  0.160 &  0.118 &  0.278 &  0.241 &  0.401 &  0.125 & 0.015 & 0.002 & 0.038 &-0.088 \nl
   0.400 &  0.225 &  0.161 &  0.388 &  0.323 &  0.549 &  0.165 & 0.039 & 0.010 & 0.050 &-0.116 \nl
   0.500 &  0.293 &  0.205 &  0.501 &  0.408 &  0.703 &  0.206 & 0.066 & 0.023 & 0.062 &-0.145 \nl
   0.600 &  0.364 &  0.250 &  0.616 &  0.492 &  0.859 &  0.246 & 0.093 & 0.038 & 0.074 &-0.173 \nl
   0.700 &  0.436 &  0.294 &  0.729 &  0.574 &  1.014 &  0.285 & 0.117 & 0.055 & 0.087 &-0.199 \nl
   0.800 &  0.510 &  0.334 &  0.842 &  0.652 &  1.165 &  0.322 & 0.141 & 0.070 & 0.099 &-0.222 \nl
   0.900 &  0.584 &  0.372 &  0.954 &  0.723 &  1.310 &  0.355 & 0.164 & 0.085 & 0.112 &-0.243 \nl
   1.000 &  0.658 &  0.407 &  1.064 &  0.791 &  1.452 &  0.387 & 0.186 & 0.100 & 0.125 &-0.262 \nl
   1.100 &  0.734 &  0.441 &  1.174 &  0.855 &  1.593 &  0.416 & 0.209 & 0.116 & 0.137 &-0.279 \nl
   1.200 &  0.812 &  0.475 &  1.285 &  0.919 &  1.733 &  0.444 & 0.234 & 0.134 & 0.149 &-0.294 \nl
   1.300 &  0.890 &  0.510 &  1.399 &  0.982 &  1.874 &  0.472 & 0.261 & 0.154 & 0.161 &-0.309 \nl
   1.400 &  0.970 &  0.548 &  1.517 &  1.047 &  2.017 &  0.499 & 0.292 & 0.179 & 0.174 &-0.324 \nl
   1.500 &  1.050 &  0.587 &  1.638 &  1.114 &  2.163 &  0.527 & 0.327 & 0.207 & 0.187 &-0.340 \nl
   1.600 &  1.128 &  0.628 &  1.759 &  1.186 &  2.312 &  0.558 & 0.362 & 0.240 & 0.202 &-0.356 \nl
   1.700 &  1.204 &  0.671 &  1.877 &  1.263 &  2.464 &  0.593 & 0.396 & 0.278 & 0.220 &-0.375 \nl
   1.800 &  1.274 &  0.714 &  1.989 &  1.347 &  2.619 &  0.634 & 0.427 & 0.323 & 0.241 &-0.396 \nl
   1.900 &  1.336 &  0.758 &  2.093 &  1.440 &  2.775 &  0.683 & 0.455 & 0.373 & 0.265 &-0.420 \nl
   2.000 &  1.390 &  0.801 &  2.189 &  1.542 &  2.933 &  0.741 & 0.482 & 0.431 & 0.295 &-0.448 \nl
   2.100 &  1.435 &  0.844 &  2.278 &  1.653 &  3.093 &  0.808 & 0.507 & 0.495 & 0.329 &-0.479 \nl
   2.200 &  1.472 &  0.888 &  2.358 &  1.773 &  3.255 &  0.883 & 0.534 & 0.565 & 0.369 &-0.512 \nl
   2.300 &  1.503 &  0.932 &  2.431 &  1.898 &  3.418 &  0.964 & 0.563 & 0.642 & 0.413 &-0.547 \nl
   2.400 &  1.530 &  0.976 &  2.502 &  2.027 &  3.581 &  1.049 & 0.594 & 0.725 & 0.461 &-0.581 \nl
   2.500 &  1.555 &  1.022 &  2.573 &  2.158 &  3.746 &  1.135 & 0.627 & 0.812 & 0.514 &-0.615 \nl
   2.600 &  1.578 &  1.069 &  2.648 &  2.290 &  3.910 &  1.222 & 0.665 & 0.905 & 0.571 &-0.645 \nl
   2.700 &  1.601 &  1.118 &  2.726 &  2.422 &  4.073 &  1.307 & 0.705 & 1.003 & 0.632 &-0.672 \nl
   2.800 &  1.623 &  1.168 &  2.802 &  2.556 &  4.232 &  1.392 & 0.742 & 1.107 & 0.697 &-0.695 \nl
   2.900 &  1.644 &  1.218 &  2.877 &  2.690 &  4.387 &  1.476 & 0.776 & 1.214 & 0.765 &-0.716 \nl
   3.000 &  1.664 &  1.270 &  2.942 &  2.827 &  4.532 &  1.558 & 0.800 & 1.330 & 0.835 &-0.730 \nl
   3.100 &  1.682 &  1.322 &  2.997 &  2.967 &  4.665 &  1.639 & 0.814 & 1.452 & 0.908 &-0.738 \nl
   3.200 &  1.697 &  1.374 &  3.042 &  3.112 &  4.782 &  1.718 & 0.821 & 1.582 & 0.983 &-0.740 \nl
   3.300 &  1.709 &  1.428 &  3.080 &  3.261 &  4.885 &  1.794 & 0.825 & 1.719 & 1.058 &-0.736 \nl
   3.400 &  1.721 &  1.483 &  3.115 &  3.409 &  4.985 &  1.868 & 0.831 & 1.859 & 1.133 &-0.730 \nl
   3.500 &  1.730 &  1.545 &  3.145 &  3.558 &  5.078 &  1.936 & 0.840 & 2.003 & 1.208 &-0.720 \nl
   3.600 &  1.738 &  1.614 &  3.181 &  3.701 &  5.178 &  1.998 & 0.860 & 2.148 & 1.283 &-0.708 \nl
   3.700 &  1.745 &  1.691 &  3.221 &  3.839 &  5.282 &  2.055 & 0.889 & 2.292 & 1.357 &-0.695 \nl
   3.800 &  1.752 &  1.774 &  3.270 &  3.973 &  5.395 &  2.107 & 0.926 & 2.436 & 1.432 &-0.682 \nl
   3.900 &  1.758 &  1.872 &  3.338 &  4.096 &  5.529 &  2.149 & 0.986 & 2.578 & 1.507 &-0.670 \nl
   4.000 &  1.765 &  1.971 &  3.408 &  4.218 &  5.666 &  2.191 & 1.047 & 2.720 & 1.583 &-0.658 \nl
   4.100 &  1.770 &  2.078 &  3.498 &  4.342 &  5.820 &  2.235 & 1.122 & 2.876 & 1.662 &-0.636 \nl
   4.200 &  1.774 &  2.185 &  3.598 &  4.474 &  5.982 &  2.287 & 1.201 & 3.047 & 1.744 &-0.605 \nl
   4.300 &  1.779 &  2.291 &  3.698 &  4.607 &  6.144 &  2.339 & 1.280 & 3.219 & 1.826 &-0.574 \nl
   4.400 &  1.783 &  2.395 &  3.805 &  4.748 &  6.308 &  2.398 & 1.369 & 3.402 & 1.910 &-0.535 \nl
   4.500 &  1.787 &  2.490 &  3.932 &  4.913 &  6.478 &  2.477 & 1.485 & 3.620 & 2.003 &-0.477 \nl
   4.600 &  1.794 &  2.599 &  4.099 &  5.091 &  6.689 &  2.546 & 1.647 & 3.850 & 2.110 &-0.419 \nl
   4.700 &  1.805 &  2.724 &  4.313 &  5.285 &  6.948 &  2.604 & 1.862 & 4.096 & 2.232 &-0.362 \nl
   4.800 &  1.816 &  2.848 &  4.527 &  5.479 &  7.207 &  2.663 & 2.077 & 4.342 & 2.354 &-0.305 \nl
   4.900 &  1.828 &  2.973 &  4.741 &  5.673 &  7.466 &  2.721 & 2.292 & 4.587 & 2.477 &-0.247 \nl
   5.000 &  1.839 &  3.098 &  4.956 &  5.867 &  7.725 &  2.779 & 2.507 & 4.833 & 2.599 &-0.190 \nl
\enddata
\end{deluxetable}

\end{document}